\documentclass{article}
\usepackage{fullpage}

\usepackage{xspace}
\usepackage{amsmath, amssymb, amsthm}

\newcommand{\push}{\textsc{push}\xspace}
\newcommand{\pop}{\textsc{pop}\xspace}

\newcommand{\enqueue}{\textsc{enqueue}\xspace}
\newcommand{\dequeue}{\textsc{dequeue}\xspace}
\newcommand{\op}{\textrm{op}}
\newcommand{\D}{\mathcal{D}}
\newcommand{\Tree}{\mathcal{T}}
\newcommand{\depth}{\textrm{depth}}
\newcommand{\E}{\mathbb{E}}
\newcommand{\supp}{\textrm{supp}}
\newcommand{\eps}{\varepsilon}

\newcommand{\poly}{\textrm{poly}}
\newcommand{\polylog}{\textrm{poly-log}}

\renewcommand{\log}{\lg}
\newtheorem{theorem}{Theorem}
\newtheorem{lemma}{Lemma}
\newtheorem{definition}{Definition}

\title{\Large Lower Bounds for Oblivious Data Structures}

\author{Riko Jacob\thanks{\texttt{rikj@itu.dk}. IT University Copenhagen.} \and Kasper Green Larsen\thanks{\texttt{larsen@cs.au.dk}. Aarhus University. Supported by a Villum Young
    Investigator grant 13163 and an AUFF starting grant.} \and  Jesper Buus Nielsen\thanks{\texttt{jbn@cs.au.dk}. Aarhus University. 
    Partially supported by the BETHE project under Independent Research Fund Denmark.
    Partially supported by the Concordium Blockchain Research Center, Aarhus University, Denmark.
    Partially supported by the European Union's Horizon 2020 research and innovation programme under grant agreement \#731583 (SODA).}}

\begin{document}
\date{}
\maketitle

\begin{abstract}
An oblivious data structure is a data structure where the memory
access patterns reveals no information about the operations performed
on it. Such data structures were introduced by Wang \emph{et al.} [ACM SIGSAC'14] and are intended for
situations where one wishes to store the data structure at an
untrusted server. One way to obtain an oblivious data structure is
simply to run a classic data structure on an oblivious RAM
(ORAM). Until very recently, this resulted in an overhead of
$\omega(\lg n)$ for the most natural setting of parameters. Moreover,
a recent lower bound for ORAMs by Larsen and Nielsen [CRYPTO'18] 
show that they always incur an overhead
of at least $\Omega(\lg n)$ if used in a black box manner. To circumvent the
$\omega(\lg n)$ overhead, researchers have instead studied
classic data structure problems more directly and have obtained
efficient solutions for many such problems such as stacks, queues, deques,
priority queues and search trees. However, none of these data
structures process operations faster than $\Theta(\lg n)$, leaving open the
question of whether even faster solutions exist. In this paper, we
rule out this possibility by proving $\Omega(\lg n)$ lower bounds for
oblivious stacks, queues, deques, priority queues and search trees.
\end{abstract}

\section{Introduction}

A number of problem domains have emerged where an algorithm needs to store data in an untrusted memory, for instance in cloud computing and trusted computing platforms like the SGX. In these settings it is typically not enough to just encrypt the data, as the access pattern itself may leak sensitive information. An \emph{oblivious data structure}\cite{DBLP:conf/ccs/WangNLCSSH14} (ODS) is one which mitigates this problem by design. An ODS admits a set of operations. On each operation it will make some accesses to the untrusted memory. The memory accesses should look independent of the operations being executed. 

In a bit more detail, security of an ODS is defined as follows. There is an adversary which picks two sequences of operations of the same length. According to a fair coinflip, one of the sequences are executed and the adversary gets to see the memory accesses made by the ODS. It then has to guess which sequence of operations was executed. The data structure is called oblivious if no adversary can guess correctly with probability significantly bounded away from one-half. 

The canonical oblivious data structure is the \emph{oblivious RAM} (ORAM), which is just an ODS for the array data structure. Given an ORAM it is much easier to construct an ODS for another data structure problem~$P$: First construct a data structure for $P$ which uses the same number of memory accesses per operation. Then run this data structure on top of the ORAM to hide the position of those memory accesses.  

The ORAM was first introduced by Goldreich and Ostrovsky \cite{DBLP:journals/jacm/GoldreichO96}. In their model the maintained array has size $n$, the server memory has size $\poly(n)$, and the word size of both the array and the untrusted memory is $\Theta(\log n)$. 
We call this the \emph{natural setting} below. In the Goldreich-Ostrovsky model, 
the client has memory size $\Theta(\log n)$ and the memory accesses leak their position to the adversary, but not the content being read or written. The rational behind leaking only positions and not contents is that the contents could simply be encrypted under a key stored by the client. Most likely due to the emerging practical applications, ORAMs have received a lot of research attention recently, see for instance \cite{DBLP:conf/crypto/PinkasR10,DBLP:conf/tcc/DamgardMN11,DBLP:conf/icalp/GoodrichM11,DBLP:conf/soda/GoodrichMOT12,DBLP:conf/soda/KushilevitzLO12,DBLP:conf/ccs/WilliamsST12,DBLP:conf/ndss/StefanovS13,DBLP:conf/asiacrypt/ChungLP14,DBLP:conf/eurocrypt/GentryHLO0W14,DBLP:conf/focs/GargLO15,DBLP:conf/tcc/BoyleCP16,DBLP:conf/crypto/LuO17,DBLP:conf/wpes/Goodrich17,Goodrich18}. 

The ORAM construction in \cite{DBLP:journals/jacm/GoldreichO96} had a bandwidth overhead of $\polylog(n)$, where the bandwidth overhead is 
defined to be the number of bits communicated with the memory divided with the number of bits read and written to the array maintained by the ORAM. The same paper also showed a lower bound on the bandwidth overhead of $\Omega(\lg n)$. The lower bound was proven in a model, where the ORAM construction is not allowed to read the data it is storing (the so-called ``balls in bins'' model) and where the adversary was allowed to have an unbounded running time.  In \cite{LN18} Larsen and Nielsen proved the same lower bound in a model where the encoding of the data can be arbitrary and where the adversary's running time is only allowed to be polynomial time. Notice that this is the more challenging model when proving a lower bound. 

The first ORAM construction to meet the lower bound for non-trivial parameters was the Path ORAM~\cite{DBLP:conf/ccs/StefanovDSFRYD13}. It has amortized bandwidth overhead of $O(\log n)$ for maintaining an array with entry size 
$\Omega(\log^2 n)$ bits using an untrusted memory with word size $\Theta(\log n)$. 
The PathORAM having $O(\lg n)$ bandwidth overhead means that it accesses $\Theta(\lg^2n)$ server cells, each storing $\Theta(\lg n)$ bits, in order to retrieve an array entry of $\Theta(\lg^2n)$ bits.
In the more natural setting where both the array and the server memory has word size $\Theta(\lg n)$ bits, the bandwidth overhead of the PathORAM is $\Theta(\lg^2 n)$. If we allow a larger client memory of $m=n^\eps$, then \cite{DBLP:journals/corr/abs-1007-1259} gave an ORAM with bandwidth overhead $O(\lg n)$. This also matches the lower bound of Larsen and Nielsen.

The PathORAM paper and the lower bound of Larsen and Nielsen leave two intriguing open problems. 1) What is the bandwidth overhead in the natural setting where both the array and the server memory has word size $\Theta(\lg n)$ bits. 2) Can the $\Theta(\lg n)$ overhead be beaten by considering data structures less powerful than the array, such as queues, stacks, deques and priority queues? There does not appear to be any reduction from the array maintenance problem to any of these, as they lack support for random access to items stored in them. Hence it is completely plausible that there are more efficient ways to solve these problems than to take the classic solutions and run them through an optimal ORAM. If so, it would be very valuable to have a library of efficient data structures for use in the oblivious setting.

Significant progress on the first question has been made recently in \cite{DBLP:journals/iacr/PatelPRY18}, where the authors present an ORAM, called PanORAMa, which achieves a bandwidth of  
 $\Theta(\lg n \cdot \lg\lg n)$ in the natural setting. After the submission of this paper, the paper~\cite{cryptoeprint:2018:892} completely solves the first problem by presenting an optimal ORAM, called OptORAMa, achieving a bandwidth overhead of $\Theta(\lg n)$ in the natural setting.

Previous work on oblivious data structures~\cite{DBLP:conf/ccs/WangNLCSSH14} suggested that there might be an affirmative answer to the second question. In~\cite{DBLP:conf/ccs/WangNLCSSH14}, the authors presented stacks, queues and deques with $O(\lg n)$ bandwidth overhead when the server cell size, block size and the client memory are all $\Theta(\lg n)$ bits. At the time these were developed, the best known ORAM construction in the natural setting had an overhead of $O(\lg^2 n)$. Thus it could appear that efficient oblivious data structures are easier to design than full generality ORAMs. It is an intriguing and important open problem whether these can be improved even further.

\subsection{Our Results}
Our main results show that there is no hope of designing more efficient oblivious data structures for the classic problems mentioned above:
\begin{theorem}
\label{thm:maininformal}[Informal]
Let $\D$ be either an oblivious stack, queue, deque, priority queue or search tree, storing $r$-bit items for $r \geq 1$ and let $n$ be an upper bound on the number of items in $\D$. If $w$ denotes the cell size of $\D$, $m$ denotes the client memory size and $\D$ has failure probability at most $1/3$, then $\D$ must have expected amortized running time $\Omega(\lg(nr/m)r/w)$. 
\end{theorem}
For the most natural setting where the cell size $w$ and the number of bits in the items, $r$, are within constant factors of each other, $w = \Theta(r)$, and for any client memory size $m \leq n^{1-\eps}$, this lower bound simplifies to $\Omega(\lg n)$. 

Since stacks, queues and deques have $O(1)$ time solutions without any requirements of being oblivious, this means that the overhead resulting from being oblivious is $\Omega(\lg n)$. For priority queues with $r,w=\Theta(\lg n)$, the best known classic solution is the priority queue resulting from Thorup's~\cite{thorup2007equivalence} reduction to sorting combined with Radix Sort, resulting in $O(1)$ amortized time per operations. For $r,w = \omega(\lg n)$ and $r=\Theta(w)$, one obtains the best classic priority queue also using Thorup's reduction to sorting combined with Han and Thorup's~\cite{han2002integer} integer sorting algorithm, resulting in a priority queue with $O(\sqrt{\lg \lg n})$ expected amortized time per operation. Thus for priority queues, being oblivious comes at a cost of at least $\Omega(\lg n/\sqrt{\lg \lg n})$. Finally, for search trees, there is the classic Fusion tree~\cite{FREDMAN1993424} supporting operations in $O(\lg_w n)$ time and the y-fast trie~\cite{WILLARD198381} supporting operations in $O(\lg w)$ time when $r = \Theta(w)$. Thus one can always implement a search tree with $O(\min\{\lg_w n, \lg w\}) = O(\sqrt{\lg n})$ time operations, hence obliviousness comes at a cost of at least $\Omega(\sqrt{\lg n})$ for search trees. We remark that the lower bound for search trees follows directly from the work of Larsen and Nielsen since search trees can solve the array maintenance problem without overhead. For the remaining problems, we need completely new proofs.

\subsection{Related Results}
Proving lower bounds for data structures has a long history, starting with the introduction of the cell probe model by Yao~\cite{Yao81}. Since then, many exciting techniques have been introduced~\cite{FS89,Miltersen:polyn,PD04a,PD06,milt:asym,patrascu06pred,PTW10,PT11,Larsen12a,Larsen12b,Yu15h,CGL15,larsenWY,weinstein:lbs}, with the current strongest lower bounds peaking at $\Omega((\lg n/\lg \lg n)^2)$~\cite{Larsen12a}. Here we briefly survey the known lower bounds related to non-oblivious priority queues and search trees (stacks, queues and deques have constant time solutions without obliviousness).

Tight lower bounds for search trees (predecessor search) were proven by P\v{a}tra\c{s}cu and Thorup~\cite{patrascu06pred} using an extension of the round elimination technique of Miltersen et al.~\cite{milt:asym}. Their lower bound holds even in the static case where no updates to the data are allowed.

For priority queues, one cannot hope to prove an $\omega(1)$ lower bound in the cell probe model when $w = \Theta(r)$  (this is disappointing since the classic upper bound is $O(\sqrt{\lg \lg n})$). The reason we cannot hope to prove $\omega(1)$ lower bounds is Thorup's reduction from priority queues to sorting: sorting can be done in linear time in the cell probe model (read the input and write it in sorted order). Hence priority queues have $O(1)$ time cell probe solutions when $w = \Theta(r)$. In the external memory model (which corresponds to $w = \omega(r)$ and were we only charge for memory accesses and not computation time), we have a non-trivial lower bound of $\Omega((r/w)\lg w/\lg \lg n)$ for priority queues that support DecreaseKeys~\cite{MR3678253} and a near-matching upper bound of $O((r/w)\lg n/\lg \lg n)$~\cite{2018arXiv180607598J}. Without DecreaseKeys, there are external memory priority queues supporting operations in $O((r/w)\lg n/\lg(m/w))$ memory accesses~\cite{MR1693788}.
\section{Lower Bounds}
We prove our lower bounds in the oblivious cell probe model of Larsen and Nielsen~\cite{LN18}. In this model, a data structure consists of a server memory of $w$-bit cells, each having an integer address in $[K]$ for some $K \leq 2^w$. A data structure is furthermore equipped with a client memory of $m$ bits, which is free to access. An oblivious cell probe data structure processes queries and updates by reading and writing to memory cells. For queries, the data structure terminates by announcing the answer to the query based on what it has probed. We refer to the reading and writing of a memory cell simply as probing it. The query and update time is defined as the number of cells it probes when processing a query and an update respectively. Randomized oblivious cell probe data structures furthermore have access to an arbitrarily long uniform random bit string $R$, which is referred to as the random oracle bit string. The bit string $R$ is drawn before any operations are performed on the data structure and is chosen independently of the future operations. We say that a randomized oblivious data structure has failure probability $\delta$ if for every sequence of operations $\op_1,\dots,\op_M$, and for every query $\op_i$ in that sequence, the probability that $\op_i$ is answered correctly is at least $1-\delta$.

When processing updates and queries, the cells probed and the contents written to cells in each step may be an arbitrary deterministic function of the client memory, random oracle bit string and contents of all other cells probed so far while processing the current operation. The data structure is also allowed to update the client memory in each step, again setting the contents to an arbitrary deterministic function of the current memory, random oracle bit string and contents of cells probed so far. Allowing an arbitrary deterministic function abstracts away the instruction set of a normal RAM and allows arbitrary computations free of charge.

To define the security requirements of an oblivious data structure, let 
$
y := (\op_1,\dots,\op_M)
$
denote a sequence of $M$ operations to a data structure problem. Let 
$$
A(y) := (A(\op_1),\dots,A(\op_M))
$$
denote the corresponding \emph{probe sequence}, where each $A(\op_i)$ is the list of probes made while processing $\op_i$. Note that $A(y)$ is a deterministic function of the random oracle bit string and the sequence $y$.
A data structure is then said to be oblivious if it satisfies the following security guarantee:
\begin{definition}[Security]\label{security}
An oblivious cell probe data structure is said to be secure if the following two properties hold:
\begin{description}
\item[Indistinguishability:]

	For any two sequences of operations $y$ and $z$ of the same length $M$, 
	their probe sequences $A(y)$ and $A(z)$ cannot be distinguished with probability better than $\frac14$ 
	by an algorithm which is polynomial time in $M$, $w$ and the logarithm of the number of different updates and queries that can be performed on it. Formally, if $\mathcal{A}_n$ denotes the image of $A$ on sequences of length $n$ and $f : \mathcal{A}_n \to \{0,1\}$ denotes a polynomial time computable function, then it must be the case that $|\Pr[f(A(y))=1] - \Pr[f(A(z))=1] | \leq \frac14$ for any two sequences $y$ and $z$ of length $n$. Here the probability is taken over the randomness of the data structure.

\item[Correctness:]

	The oblivious cell probe data structure has failure probability at most $1/3$ per operation. 

\end{description}
\end{definition}
We remark that the above definition of indistinguishability requires that the adversary is deterministic. Since we are proving lower bounds, this restriction only strengthens our results, i.e., even if one needs only be secure against deterministic polynomial time adversaries, then our lower bounds still hold.
With these definitions, we are ready to proceed to our proofs. We refer readers interested in a more formal definition of oblivious cell probe data structures to~\cite{LN18}.

\subsection{Stacks}
In this section, we prove our lower bound for stacks and conclude by describing how a small change in the argument yields a similar lower bound for queues. Recall that a stack supports the two operations \push and \pop, where \push adds an $r$-bit value to the ``top'' of the stack, and \pop removes and returns the element last pushed to the stack (the ``top'' element). We prove the following lower bound:

\begin{theorem}
\label{thm:mainstack}
Let $\D$ be a stack implemented in the oblivious cell probe model, with $r$-bit elements for $r \geq 1$ and let $n \geq C$ be an upper bound on the number of elements in $\D$, where $C >1$ is some universal constant. If $w \geq 1$ denotes the cell size of $\D$, $m \geq w$ denotes the client memory size and $\D$ has failure probability at most $1/3$, then there exists a sequence $y$ of $n$ operations such that $\D$ must have expected amortized running time $\Omega(\lg(nr/m)r/w)$ on $y$. 
\end{theorem}

We let $\D$ be a stack as in Theorem~\ref{thm:mainstack}, except that we assume it has failure probability no more than $1/32$. Note that one can always obtain a stack with failure probability $1/32$ by running a constant number of independent copies in parallel of a stack with failure probability $1/3$. On a \pop operation, we simply use a majority vote to determine the result. This increases the running time and memory size by only a constant factor. Throughout the proof, we let $[K] \subseteq [2^w]$ denote the set of possible addresses of the memory cells of $\D$.

To prove the theorem, we define a hard/expensive sequence $y$ of $n$ operations. The sequence is as follows:
$$
y := \push(\bar{0}),\pop, \push(\bar{0}),\pop,\dots,\push(\bar{0}),\pop
$$
where $\bar{0}$ is the all-zeroes bit string of length $r$. The sequence $y$ thus repeatedly pushes one element onto the stack, namely the all-zeroes bit string, and then immediately pops it again. This is done for a total of $n$ times, i.e., $n/2$ \push operations and $n/2$ \pop operations. The maximum size of the stack at any given time is thus $1$. What makes $y$ expensive in the oblivious setting, is that $\D$'s processing of $y$ must be indistinguishable from many other sequences of $n$ operations.

To prove that $y$ is expensive, we use the approach of Larsen and Nielsen~\cite{LN18} in their lower bound proof for online ORAMs. Their approach is an extension of the earlier \emph{information transfer} method of 
P\v{a}tra\c{s}cu and Demaine~\cite{PD06}, modified to capture the obliviousness requirement. The setup is as follows: Define an information transfer tree $\Tree$ as a binary tree with exactly $n$ leaves. We assume that $n$ is a power of two such that $\Tree$ is a perfect binary tree. For any sequence $z = \op_1,\dots,\op_n$ of $n$ operations, associate the $i$'th operation $\op_i$ in $z$ with the $i$'th leaf of $\Tree$. When $\D$ processes the sequence $z$, it performs the probes $A(z) = (A(\op_1),\dots,A(\op_n))$ where each $A(\op_i)$ is a sequence of probes $p_{i,1},\dots,p_{i,k_i}$ performed by $\D$ when processing the operation $\op_i$. We assign every probe $p_{i,j}$ in every $A(\op_i)$ to a node in $\Tree$ as follows: For a probe $p_{i,j}$, let $s \in [K] \subseteq [2^w]$ denote the address of the memory cell probed. Let $p_{i',j'}$ be the last probe prior to $p_{i,j}$ which also probed the cell with address $s$. We assign $p_{i,j}$ to the lowest common ancestor of the two leaves corresponding to $\op_{i'}$ and $\op_{i}$ (leaf number $i'$ and $i$). If $p_{i',j'}$ does not exist, i.e., the cell with address $s$ was never probed before, then we do not assign $p_{i,j}$ to any node in $\Tree$. 

Our goal is to show that for the sequence $y$, there must be many nodes in $\Tree$ that have many probes assigned to them. Since a probe is only assigned to one node in $\Tree$, this implies that there must be a large number of probes in $A(y)$. To prove this, define $\depth(v)$ for a node $v \in \Tree$ to be the distance from $v$ to the root of $\Tree$, i.e., the root has depth $0$ and the leaves have depth $\lg n$. If we use $P_v(z)$ to denote the set of probes assigned to a node $v \in \Tree$ for a sequence of $n$ operations $z$, then we prove the following:
\begin{lemma}
\label{lem:nodemany}
If $\D$ has failure probability at most $1/32$  and $n \geq C$ for some universal constant $C>1$, then for every internal node $v \in \Tree$ of depth $d  = \depth(v) \in \{5,\dots,(1/2)\lg(nr/m)\}$, it must be the case that
$$
\E[|P_v(y)|] = \Omega(nr/(w2^d)).
$$
\end{lemma}
Intuitively, the lemma says that $P_v(y)$ contains $\Omega(r/w)$ probes \emph{for every operation} in $v$'s subtree. Now, since an operation is contained in the subtree rooted at $\Omega(\lg(nr/m))$ nodes with a depth in $\{5,\dots,\lg(nr/m)\}$, it follows that every operation must incur $\Omega(\lg(nr/m)r/w)$ probes in total.
Before proving Lemma~\ref{lem:nodemany}, we prove formally that it implies Theorem~\ref{thm:mainstack}. For this, observe that by linearity of expectation and the fact that any probe in $A(y)$ is assigned to at most one node of $\Tree$, we have:
\begin{eqnarray*}
\E[|A(y)|] &\geq& \sum_{v \in \Tree(y)} \E[|P_v(y)|] \\
&\geq& \sum_{d = 5}^{(1/2)\lg(nr/w)} \sum_{v \in \Tree(y) : \depth(v) = d} \E[|P_v(y)|] \\
&=& \Omega\left( \sum_{d = 5}^{(1/2)\lg(nr/w)} 2^d \cdot (nr/(w2^d)) \right) \\
&=& \Omega(n\lg(nr/w)r/w).
\end{eqnarray*}
Since there are $n$ operations in $y$, this implies that the expected amortized running time of $\D$ is $\Omega(\lg(nr/w)r/w)$. 

We thus set out to prove Lemma~\ref{lem:nodemany}. For this, let $v \in \Tree$ be a node with $d=\depth(v) \in \{5,\dots,(1/2)\lg(nr/m)\}$. Consider the following random sequence $Z_v = \op_1, \dots,\op_n$ of $n$ operations: For every operation $\op_i$ not associated to a leaf in $v$'s subtree, we let $\op_i = \push(\bar{0})$ if $i$ is odd and we let $\op_i=\pop$ if $i$ is even. Thus $Z_v$ is identical to the sequence $y$ outside the subtree rooted at $v$. The $n/2^{d+1}$ operations $\op_j,\dots,\op_{j + n/2^{d+1}}$ in $v$'s left subtree are $\push(b_1),\dots,\push(b_{n/2^{d+1}})$ where each $b_i$ is a uniform random and independently chosen $r$-bit string. The $n/2^{d+1}$ operations in $v$'s right subtree are all $\pop$ operations. Thus for the sequence $Z_v$, we start by repeatedly pushing and popping $\bar{0}$ until we reach $v$'s left subtree. We then push $n/2^{d+1}$ random bit strings onto the stack and pop them again in $v$'s right subtree. The intuition is that while $\D$ processes the \pop operations in $v$'s right subtree, it needs to recover all the random bits that were pushed during the operations in $v$'s left subtree. By an entropy argument, this means that $\D$ must read $\Omega(nr/2^{d+1})$ bits containing information about $b_1,\dots,b_{n/2^{d+1}}$. This costs $\Omega(nr/(2^{d+1}w))$ probes as cells have $w$ bits. Now observe that those probes must be assigned to $v$. To see this, note that if they were assigned to ancestors of $v$, then they read information that was written prior to the operations in $v$'s left subtree were performed. Hence they cannot reveal useful information about $b_1,\dots,b_{n/2^{d+1}}$. Similarly, if they were assigned to descendants of $v$, then they must be assigned to nodes in $v$'s right subtree. But this means that the cell was already probed during $v$'s right subtree and thus the probe reveals no new information about $b_1,\dots,b_{n/2^{d+1}}$. We will use this line of argument to show that:
\begin{lemma}
\label{lem:readunderZ}
If $\D$ has failure probability at most $1/32$ and $n \geq C$ for some universal constant $C>1$, then for every internal node $v \in \Tree$ of depth $d  = \depth(v) \in \{5,\dots,(1/2)\lg(nr/m)\}$, it must be the case that
$$
\Pr[|P_v(Z_v)| \geq (1/100)nr/(w2^d)] \geq 1/2.
$$
\end{lemma}
Lemma~\ref{lem:readunderZ} shows that for the random sequence $Z_v$, there must often be many probes assigned to $v$. Before proving Lemma~\ref{lem:readunderZ}, we use it to finish the proof of Lemma~\ref{lem:nodemany}. The key idea is that the indistinguishability assumption forces the behaviour of $\D$ to look identical regardless of whether it is processing $y$ or $Z_v$. Thus in some sense $|P_v(Z_v)|$ and $|P_v(y)|$ have to be similar, i.e., $y$ must also force many probes to be assigned to $v$. We argue formally as follows: 
Assume for the sake of contradiction that $\E[|P_v(y)|] \leq (1/400)nr/(w2^d)$. Then by Markov's inequality, we have $\Pr[|P_v(y)| \geq (1/100)nr/(w2^d)] \leq 1/4$. Moreover,  Lemma~\ref{lem:readunderZ} and an averaging argument implies that there exists a sequence $z \in \supp(Z_v)$ such that $\Pr[|P_v(z)| \geq (1/100)nr/(w2^d)] \geq 1/2$. The following procedure now distinguishes $y$ and $z$ with probability at least $1/4$: Let $a \in \{y,z\}$ be one of the two sequences of $n$ operations. From $A(a)$, compute the set $P_v(a)$. If $|P_v(a)| \geq (1/100)nr/(w2^d)$ output $1$. Otherwise output $0$. Notice that for $a = z$, this algorithm outputs $1$ with probability at least $1/2$, whereas if $a = y$, it outputs $1$ with probability at most $1/4$, i.e., the algorithm distinguishes $y$ and $z$ with probability at least $1/4$.  Since the algorithm uses only $A(a)$ to construct $P_v(a)$ and since the algorithm runs in time polynomial in $n$, this contradicts $\D$ being an oblivious stack and completes the proof of Lemma~\ref{lem:nodemany}.

What remains is to prove Lemma~\ref{lem:readunderZ}. As mentioned above, our proof will exploit that the \pop operations in $v$'s right subtree have to recover all the random bits inserted into the stack by the \push operations in $v$'s left subtree. The formal proof proceeds via an encoding argument.  We have two players Alice (encoder) and Bob (decoder) that share the random oracle bit string $R$ used by $\D$. Alice is given $b_1,\dots,b_{n/2^{d+1}}$ as input and must send a message to Bob such that Bob can recover $b_1,\dots,b_{n/2^{d+1}}$ from the message and $R$. Let $H(\cdot )$ denote binary entropy. Since $b_1,\dots,b_{n/2^{d+1}}$ are uniform random $r$-bit strings that are chosen independently of each other and the random oracle bit string $R$, we have 
$$
H(b_1 \dots b_{n/2^{d+1}} \mid R) = nr/2^{d+1}.
$$
It follows from Shannon's source coding theorem than any encoding (message of Alice) must have an expected length of at least $nr/2^{d+1}$ bits if it allows Bob to recover $b_1,\dots,b_{n/2^{d+1}}$. We now assume for the sake of contradiction Lemma~\ref{lem:readunderZ} is false, i.e., $\D$ has failure probability at most $1/32$ and satisfies $\Pr[|P_v(Z_v)| \geq (1/100)nr/(w2^d)] < 1/2$. Our goal is to show that this assumption allows Alice to send a message shorter than $nr/2^{d+1}$ bits in expectation while still allowing Bob to recover $b_1,\dots,b_{n/2^{d+1}}$. This is an information theoretic contradiction.

\paragraph{Encoding.} Alice is given $b_1,\dots,b_{n/2^{d+1}}$ and $R$. She proceeds as follows:
\begin{enumerate}
\item She starts by constructing the sequence $Z_v$ from $b_1,\dots,b_{n/2^{d+1}}$. Let $Z_v^0$ denote the subsequence consisting of all operations preceeding $v$'s subtree. Let $Z_v^r$ denote the operations in $v$'s left subtree (the \push operations) and let $Z_v^\ell$ denote the operations in $v$'s right subtree (the \pop operations). Alice runs the sequence of operations $Z^0_v \circ Z^\ell_v \circ Z^r_v$ on $\D$ using the randomness $R$. Here $\circ$ denotes concatenation and $Z^0_v \circ Z^\ell_v \circ Z^r_v$ is simply the sequence of operations up to and including $v$'s subtree. While doing so, she computes the set $P_v(Z_v)$ (this set does not depend on operations after $v$'s subtree) and the set $Q$ containing all \pop operations in $v$'s right subtree that fail to return the correct answer. If either $|P_v(Z_v)| \geq (1/100)nr/(w2^d)$ or if $|Q| \geq (1/8)n/2^{d+1}$, then Alice sends a $0$-bit followed by a naive encoding of $b_1,\dots,b_{n/2^{d+1}}$, costing $nr/2^{d+1}$ bits. Otherwise Alice sends a $1$-bit and proceeds to the next step.
\item Alice now sends Bob the contents of the client memory as it was immediately after processing $Z^0_v \circ Z^\ell_v$, costing $m$ bits. She also sends all cells probed by probes in $P_v(Z_v)$. When sending the cells probed by probes in $P_v(Z_v)$, she sends their addressess as well as their contents as they were immediately after processing $Z^0_v \circ Z^\ell_v$. Accounting for specifying $|P_v(Z_v)|$, this costs at most $\lg n + ((1/100)nr/(w2^{d+1})) \cdot (\lg K + w)$ bits. Finally, Alice also sends the set $Q$ together with the correct answer to every \pop operation in $Q$. This costs at most $\lg n +(1/8)nr/2^{d+1} + \lg \binom{n/2^{d+1}}{(1/8)n/2^{d+1}}$ bits. This concludes Alice's message.
\end{enumerate} 

\paragraph{Decoding.} Bob is given $R$ and Alice's message. His task is to recover $b_1,\dots,b_{n/2^{d+1}}$. He does as follows:
\begin{enumerate}
\item He starts by checking the first bit of Alice's message. If this is a $0$-bit, he immediately recovers $b_1,\dots,b_{n/2^{d+1}}$ from the remaining part of the message (the naive encoding). Otherwise, he proceeds to the next step.
\item Bob can privately compute $Z_v^0$ and $Z_v^r$ as they are independent of $b_1,\dots,b_{n/2^{d+1}}$. He does so and now runs $Z_v^0$ on $\D$ using the randomness $R$. Once this has finished, he updates the client memory to what Alice told him it was after $Z_v^0 \circ Z_v^\ell$. He also overwrites the contents of every cell probed in $P_v(Z_v)$ with the contents as they were after processing $Z_v^0 \circ Z_v^\ell$. Finally, he skips over all operations in $Z_v^\ell$ and runs all the operations in $Z_v^r$ on $\D$. He collects the list of answers returned by the \pop operations in $Z_v^r$ and finally corrects the answers to all those \pop operations that appear in $Q$. We claim that all \pop operations in $Z_v^r$ now give the correct answer for the sequence $Z_v^0 \circ Z_v^\ell \circ Z_v^r$ and thus Bob recovers $b_i$ from the answer to the $(n/2^{d+1}-i+1)$'th \pop operation in $Z_v^r$. To see this, observe that before applying the corrections in $Q$, all \pop operations give the same answers as when Alice executed them. This is true for the following reason: Every time a probe is made in $Z_v^r$, Bob has the same contents of the probed cell as Alice did during her processing of the operation. Indeed, consider the first time a cell with some address $s \in [K]$ is probed during the processing of $Z_v^r$. If that probe is in $P_v(Z_v)$, then the contents are correct since he updated the contents based on Alice's message. If the probe was assigned to an ancestor of $v$, it must be the case that the cell was not updated during Alice's processing of $Z_v^\ell$ (by the way we assigned probes). Hence the contents Bob has from his simulation of $Z_v^0$ are also the correct contents after processing $Z_v^0 \circ Z_v^\ell$. Finally, the cell cannot be assigned to a descendant of $v$ as this would contradict that it is the first time it is being probed during $Z_v^r$. It follows that Bob always has the same contents of the probed cells as Alice did, and hence his answers to the \pop operations are consistent with Alice's answers (they use the same randomness $R$).
\end{enumerate}

\paragraph{Analysis.}
What remains is to analyse the expected size of the encoding to derive our contradiction. Consider first the case in which Alice's message starts with a $1$-bit. In this case, Alice sends no more than
\begin{eqnarray*}
m + 2\lg n + ((1/100)nr/(w2^{d})) \cdot (\lg K + w) +
(1/8)nr/2^{d+1} + \lg \binom{n/2^{d+1}}{(1/8)n/2^{d+1}}
\end{eqnarray*}
bits. Using that $\lg K \leq w$ by assumption, $\binom{n/2^{d+1}}{(1/8)n/2^{d+1}} \leq (8e)^{(1/8)n/2^{d+1}}$ and $m \leq nr/2^{2d} = (nr/2^{d+1})/2^{d-1} \leq (nr/2^{d+1})/16$ (here we use both $d \geq 5$ and $d \leq (1/2)\lg(nr/m)$), we can bound the above by:
\begin{eqnarray*}
(1/16)nr/2^{d+1} + 2\lg n + (1/25)nr/2^{d+1} +
(1/8)nr/2^{d+1} + (\lg(8e)/8)n/2^{d+1} &\leq& \\
2 \lg n + (1/16 + 1/25 + 1/8 +56/100)nr/2^{d+1} &\leq& \\
2 \lg n + (79/100)nr/2^{d+1} &\leq& \\
(99/100)nr/2^{d+1}.
\end{eqnarray*}
In the last line, we used that $nr/2^{d+1} \geq \sqrt{nrm}/2$ which for $n$ at least some sufficiently large constant is much bigger than $2 \lg n$.
Letting $C$ denote the event that Alice's message starts with a $0$-bit, we get that the expected number of bits sent is
$$
1 + \Pr[C]nr/2^{d+1} + (1-\Pr[C])(99/100)nr/2^{d+1}.
$$ 
This is increasing in $\Pr[C]$, thus we upper bound $\Pr[C]$. First note that $\Pr[|P_v(Z_v)| \geq (1/100)nr/(w2^d)] < 1/2$ by assumption. Moreover, we have $\E[|Q|] = (1/32)n/2^{d+1}$ since $\mathcal{D}$ has failure probability at most $1/32$. Hence $\Pr[|Q| \geq (1/8)n/2^{d+1}] \leq 1/4$. By Markov's inequality and a union bound, we have $\Pr[C] \leq 1/2 + 1/4 = 3/4$. Therefore the expected number of bits sent by Alice is upper bounded by:
\begin{eqnarray*}
1 + (3/4)nr/2^{d+1} + (1/4)(99/100)nr/2^{d+1} &<&\\
 nr/2^{d+1} &=& \\
H(b_1 \cdots b_{n/2^{d+1}} \mid R)
\end{eqnarray*}
which gives our contradiction.

\subsection{Other Data Structures}
In the following, we discuss extensions of the above lower bound to other oblivious data structures. First consider a queue supporting \enqueue and \dequeue operations. Observe that we can repeat all arguments in the proof of our stack lower bound, replacing any \push operation by an \enqueue operation and every \pop operation by a \dequeue operation. The only difference is in the encoding game where Bob uses the answer from the $i$'th \dequeue operation to recover $b_i$ instead of using the $(n/2^{d+1}-i+1)$'th \pop operation (because a queue is First-In-First-Out as opposed to the Last-In-First-Out behaviour of a stack). Thus the lower bound of Theorem~\ref{thm:mainstack} also applies to queues.

Since a deque can simulate both a stack and a queue, the same lower bound naturally applies to deques as well. A priority queue can simulate a queue simply by inserting items with increasing priority, hence the same running time lower bound applies to priority queues as well. Finally, a search tree can simulate all of these, as well as an array, hence the lower bound also applies to search tree and we conclude:
\begin{theorem}
\label{thm:main}
Let $\D$ be either a stack, queue, deque, priority queue or search tree implemented in the oblivious cell probe model, with $r$-bit elements for $r \geq 1$ and let $n$ be an upper bound on the number of elements in $\D$. If $w$ denotes the cell size of $\D$, $m$ denotes the client memory size and $\D$ has failure probability at most $1/3$ per operation, then there exists a sequence $y$ of $n$ operations such that $\D$ must have expected amortized running time $\Omega(\lg(nr/m)r/w)$ on $y$. 
\end{theorem}

\bibliographystyle{abbrv}
\bibliography{refs}

\end{document}